\documentstyle[12pt]{article}
\begin{document}

\newtheorem{theorem}{Proposition}

\begin{center}
{\large{\bf General vorticity conservation}} \\
\vspace{1cm}
{\large{Hasan G\"{u}mral}}  \\  
\vspace{5mm}
T\"UB\.ITAK-Feza G\"{u}rsey Institute  \\
P.O. Box 6, 81220 \c{C}engelk\"oy-\.Istanbul, Turkey \\ 
e-mail:hasan@mam.gov.tr  \\
\vspace{5mm}
\today
\end{center}

\vspace{1cm}

\section*{Abstract}
The motion of an incompressible fluid in Lagrangian coordinates
involves infinitely many symmetries generated by the left Lie
algebra of group of volume preserving diffeomorphisms
of the three dimensional domain occupied by the fluid.
Utilizing a $1+3$-dimensional Hamiltonian setting an explicit
realization of this symmetry algebra and the related
Lagrangian and Eulerian conservation laws are constructed recursively.
Their Lie algebraic structures are inherited from the same construction.
The laws of general vorticity and helicity conservations are
formulated globally in terms of invariant differential forms of the
velocity field.

\newpage

\section{Introduction}

The configuration space of an incompressible fluid is the
group of volume preserving transformations of the three dimensional
region in $R^3$ containing the fluid.
The motion is generated by the left action of the group
by composition and hence the velocity field is right invariant.
Thus, the generators of the right action which form the infinite
dimensional left Lie algebra are infinitesimal
symmetries of the velocity field \cite{arnold}-\cite{mr}.
In the fluid mechanical context, this is known as the particle
relabelling symmetry \cite{mr},\cite{salmon},
also named as gauge transformations
in Ref.\cite{hen} and as trivial displacements in Ref.\cite{schutz}.
It has been the subject of many investigations to discover the
corresponding Lagrangian and Eulerian invariants in their
most general form and to express for them the so-called
general vorticity conservation law.
(see Ref.\cite{salmon} for a discussion and references).

One the best illustrative example for this general
conservation principle may be the equivalences of
the preservation of coadjoint orbits of
vorticity, the right invariance on the cotangent bundle of the
group of volume preserving diffeomorphisms, the Helmholtz'
vorticity and the Kelvin's circulation theorems
(see \cite{mr} or section(\ref{sgvc})).
Another formulation of the same principle results, in addition,
to the Eulerian conservation laws for helicity \cite{salmon}.
Both examples have in common the property of being derived
from the particle relabelling symmetry.

In spite of the fact that the particle relabelling
symmetry is a direct consequence of the description of motion itself,
many investigations on the structure of symmetries and
invariants employ the techniques of analysing the defining
equations for them thereby obtaining an algebraically and
quantitatively incomplete picture of symmetries, invariants
and their connections.

In this work we shall present a geometric framework for a
systematic study of symmetries and invariants which will
provide, by addressing explicitly to Lagrangian and Eulerian
formulations, a better understanding of them in the description
of motion.

\subsection{Summary and content}

We shall consider the Lagrangian description of fluid motion
in the framework of a formal geometric structure to obtain a realization
of the symmetry algebra, the corresponding Lagrangian invariants
and the general vorticity conservation law.

We shall start from the observation that the Euler equations
themselves are the conditions for the vorticity field to be
an infinitesimal symmetry of the
velocity field. We shall use this symmetry to construct
a symplectic structure on the time-extended domain in $R \times R^3$.
This will define the Hamiltonian structure of the suspended
velocity field.
Identifying a Hamiltonian vector field as a new symmetry of
the velocity field, we shall be able to generate the infinitesimal
symmetries recursively.

The vorticity is, by construction, an automorphism of the
Hamiltonian structure. Since any Hamiltonian vector field
is also an automorphism, we shall conclude that the resulting infinite
hierarchy of vector fields do form the Lie algebra of symmetries.

Associated to each infinitesimal symmetry we shall then introduce
invariant differential forms of the velocity field. Among a plethora of
differential invariants of various degrees
we shall identify a hierarchy and use it
as the basic ingredient for the global expressions of the laws of
general vorticity and helicity conservations.

After a brief description of fluid motion in the next section
we shall recall from Ref.\cite{hg97} the symplectic structure of the
suspended velocity field on the time-extended space $R \times R^3$.
In section(\ref{symmetry}), we shall obtain the time-dependent symmetries
on $R \times R^3$ of the velocity field and
show that an equivalent representation
of them are Hamiltonian vector fields on $R^3$.
In section(\ref{invariant}), we shall give explicit expressions
for the laws of general vorticity and helicity conservations.
All of these will arise in connection with the infinitesimal
automorphisms, restriction to spatial domain in $R^3$, and the associated
invariant forms of the symplectic structure.

\section{Three dimensional fluid motion}

Let the open set $D \subset R^{3}$ be the domain occupied initially
by an incompressible fluid and 
$x(t=0)=x_{0} \in D$ be the initial position, i.e., a Lagrangian label.
For a fixed initial position $x_{0}$,
the Eulerian coordinates $x(t) = g_{t}(x_{0})$ define 
a smooth curve in $R^{3}$
describing the evolution of fluid particles. 
For each time $t \in R$, the volume preserving embedding 
$g_{t}:D \to g_t(D)=D_t \subset R^{3}$ describes a
configuration of fluid. A flow is then a curve $t \mapsto g_{t}$ 
in the space of all such transformations.
The time-dependent Eulerian (spatial) velocity field $v_{t}$
that generates $g_{t}$ is defined by
\begin{equation}
    {dx \over dt}={dg_{t}(x_{0}) \over dt}
            =(v_{t} \circ g_{t})(x_{0})=v(t,x)    \label{vel}
\end{equation}
where $v_{t} \circ g_{t}$ is the corresponding Lagrangian (material)
velocity field \cite{via},\cite{mr}.
Since $g_{t}$ is volume preserving, $v_{t}(x)$ is a divergence-free vector
field over $R^{3}$ and Eq.(\ref{vel}) is a non-autonomous dynamical system
associated with it.
The system (\ref{vel}) can, equivalently, be represented as an
autonomous system defined by the suspended velocity field
\begin{equation}
  \partial_{t} + v(t,x)   \;,\;\;\;\;\; v={\bf v} \cdot \nabla \label{tvel}
\end{equation}
on the time-extended space $R \times R^{3}$.

The Lagrangian description of fluid motion is the description
by trajectories, that is, by solutions 
of non-autonomous ordinary differential equations (\ref{vel}).
Equivalence of (\ref{vel}) to the autonomous system associated
with (\ref{tvel}) means that the trajectories can be obtained
by describing streamlines at each time.
In Ref. \cite{hg97}, using the Eulerian dynamical equations,
we constructed a formal symplectic structure for (\ref{tvel})
on a time-extended domain in $R \times R^3$. We shall now summarize
this construction.

\subsection{Symplectic structure}

A symplectic structure \cite{via},\cite{mr},\cite{olver}
on a manifold $N$ of even dimension $2n$ is defined by a closed,
non-degenerate two-form $\Omega$. It is exact if there exists a 
one-form $\theta$ such that $\Omega =- d \theta$.
Darboux's theorem quarantees the existence of 
local coordinates $(q^{i},p_{i})\; i=1,...,n$ 
in which $\Omega$ has the canonical form $dq^{i} \wedge dp_{i}$.
The $2n-$form $(-1)^{n} \Omega^{n}/n!$ is called the Liouville volume.
A vector field $V$ on $N$ is called Hamiltonian if there exists
a function $h$ on $N$ such that 
\begin{equation}
      i(V)(\Omega)=dh                 \label{seq}
\end{equation}
where $i(V)(\cdot)$ denotes the inner product with $X$.
The identity $i(V)(dh)=0$ which follows from (\ref{seq}) is 
the expression for conservation of $h$ under the flow of $V$.
With the correspondence (\ref{seq}) between functions and vector fields,
the Poisson bracket of functions on $N$ defined by
\begin{equation}
   \{ f,g \} = \Omega (V_{f},V_{g}) = \Omega^{-1}(df,dg)   \label{pobi}
\end{equation}
satisfies the conditions of bilinearity, skew-symmetry, the Jacobi identity
and the Leibniz rule.
This enables us to write the dynamical system associated with the
vector field $V_{h}$ in the form of Hamilton's equations
\begin{equation}
   {dx \over dt} = \{ x,h  \}        \;.      \label{heq}
\end{equation}

\begin{theorem}
Let the dynamics of the velocity field ${\bf v}$ be governed by 
\begin{equation}
  {\partial {\bf v} \over \partial t} +
  {\bf v} \cdot \nabla {\bf v}= {\bf F}   \label{geuler}
\end{equation}
and assume that the divergence-free field ${\bf B}$ and the function
$\varphi$ satisfy
\begin{equation}
    {\partial {\bf B} \over \partial t} - 
    \nabla \times ( {\bf v} \times {\bf B} ) =0 \;,\;\;\;
    {\partial \varphi \over \partial t} +
     {\bf v} \cdot \nabla \varphi =0     \label{beq}
\end{equation}
which are the frozen field equations. Then

(1) $\partial_{t}+v$ is a Hamiltonian vector field with the symplectic
two-form
\begin{equation}
  \Omega = - (\nabla \varphi + {\bf v} \times {\bf B}) 
          \cdot d{\bf x} \wedge dt + {\bf B} \cdot 
             (d{\bf x} \wedge d{\bf x})    \label{symp2}
\end{equation}
and the Hamiltonian function $\varphi$.

(2) $\rho_{\varphi} \equiv - {\bf B} \cdot \nabla \varphi$ is the invariant
Liouville volume density. 

(3) If moreover ${\bf B} = \nabla \times {\bf A}$ for some vector 
potential ${\bf A}$, then $\Omega$ is exact
\begin{equation}
 \Omega = -d \theta \;,\;\;\; 
        - \theta =  \psi \, dt + {\bf A} \cdot d{\bf x}     \label{cone2}
\end{equation}
where $\psi$ is determined by the equation 
\begin{equation}
  {\partial {\bf A} \over \partial t} -
  {\bf v} \times (\nabla \times {\bf A}) = 
  \nabla (\varphi + \psi)     \;.            \label{aeq}
\end{equation}

(4) In Darboux coordinates,
$\Omega = dq \wedge dp + dt \wedge dh_{can}$ and the potential field
has the Clebsch representation 
\begin{equation}
  {\bf A}= \nabla s  + p \nabla q
\end{equation}
where $s$ is the generating function of the canonical transformation.
\end{theorem}

The Euler flow of an ideal isentropic fluid
is characterized by ${\bf F}= -\nabla(P/ \rho)$, where
$P$ and $\rho$ are pressure and density, respectively.
In this case, ${\bf A}$ is replaced by the velocity
field itself, ${\bf B}$ becomes the vorticity
${\bf w}=\nabla \times {\bf v}$ and Eqs.(\ref{aeq}) reduces
to the expression
\begin{equation}
    - \psi = \varphi + P / \rho +v^2 /2       \label{psio}
\end{equation}
for the scalar potential. The invariant volume density
$\rho_{\varphi} =- {\bf w} \cdot \nabla \varphi$ 
is also known as potential vorticity.

\section{Symmetries}         \label{symmetry}

A time-dependent vector field $U = \xi \partial_{t}+u$ 
is an infinitesimal geometric symmetry of dynamical system
described by $v$ if the criterion
\begin{equation}
  [ \partial_{t}+v, \xi \partial_{t}+u ] = 
                (\xi_{,t}+v(\xi))(\partial_{t}+v)      \label{symm}
\end{equation}
is satisfied. These are the most general symmetries of the system
(\ref{vel}) of first order ordinary differential equations \cite{olver}.

The frozen field equation (\ref{beq}) for ${\bf B}$ is an
expression for it to be a symmetry of $v$. Since $\rho_{\varphi}$
is a conserved function, $\rho_{\varphi}^{-1} {\bf B}$ is also
a symmetry.
We observe that this initial symmetry leaves the symplectic
two-form (\ref{symp2}) invariant. More precisely, one checks that
the Lie derivative of $\Omega$ with respect to the normalized
field $\rho^{-1}_{\varphi}{\bf B} \cdot \nabla$ vanishes.

Our intention now is to find another infinitesimal invariance
of $\Omega$ which is a symmetry of the velocity field as well.
One of the best candidates for this is a Hamiltonian vector field
because the symplectic two-form is invariant under the flows of 
Hamiltonian vector fields.
This can easily be seen from the identity
${\cal L}_{U}=i(U)d+di(U)$ for the Lie derivative
together with Eq.(\ref{seq}) and the closure of $\Omega$.
If $U$ is such a field, called an automorphism of $\Omega$,
then it follows from the identity
\begin{equation}
{\cal L}_{[ \rho^{-1}_{\varphi}B,U]}=
  {\cal L}_{ \rho^{-1}_{\varphi}B } {\cal L}_{U} -
  {\cal L}_{U} {\cal L}_{ \rho^{-1}_{\varphi}B }     \label{iden}
\end{equation}
that $[ \rho^{-1}_{\varphi}B,U]$ also leaves $\Omega$ invariant.
Replacing $ \rho^{-1}_{\varphi}B $ with $[ \rho^{-1}_{\varphi}B,U]$
in the identity (\ref{iden}) we see that one can generate an
infinite dimensional algebra of Hamiltonian vector fields
(over simply connected domains of fluid) as
invariants of the symplectic two-form.

Thus, if we can find $U$ which
is also a symmetry of the velocity field,
this so-called algebra of symplectic
automorphisms of $\Omega$ will be carried over to 
the symmetry algebra of the velocity field. The identity (\ref{iden})
will then become the Jacobi identity of the algebra of vector
fields and will enable us to obtain the generators recursively.
\begin{theorem} The Hamiltonian vector field
\begin{equation}
   U= \rho^{-1}_{\varphi} [ - B(h) (\partial_{t} +v) + {dh \over dt} B 
    + \nabla \varphi \times \nabla h  \cdot \nabla  ]  \label{fluid}
\end{equation}
associated with the symplectic two-form $\Omega$ and
the arbitrary smooth function $h$ is an infinitesimal symmetry of $v$ if 
$h$ is invariant under the flow of $v$, that is, $dh/dt=h_{,t}+v(h)=0$.
In this case, the brackets
\begin{equation}
        [...[[ \rho^{-1}_{\varphi} B,U ],U ],...]    \label{infsym}
\end{equation}
generate an infinite hierarchy of time-dependent infinitesimal
Hamiltonian symmetries of the velocity field $v$.
\end{theorem}

The infinitesimal symmetries (\ref{infsym}) are of the form of
\begin{equation}
   U_{k} = \xi_k (\partial_t +v) + \hat{u}_k  \;,\;\;\; k=0,1,2,...
\label{uk}  \end{equation}
where $\xi_k$'s are conserved functions of $v$ and $\hat{u}_k$'s
are vector fields on the spatial domain in $R^3$.
We compute $\xi_0=0 ,\; \xi_1 = -B(h) ,\; \xi_2 = - B^2(h) ,\;
\xi_3 = \hat{u}_2 (\xi_1)
- \hat{u}_1 (\xi_2),\; \xi_4 = \hat{u}_3 (\xi_1)
-\hat{u}_1 (\hat{u}_2 (\xi_1)) + \hat{u}_{1}^{2} (\xi_2) ,\; ...$
to list a few of these functions.
The vector fields $\hat{u}_k$ have a well-known interpretation
in symmetry analysis of differential equations \cite{olver}. 
They are the unique characteristic (or evolutionary) forms
of $U_k$'s along the velocity field $v$.
It follows from Eq.(\ref{symm}) by direct computation that
the symmetry condition for $\hat{u}_k$ reduces to
\begin{equation}
  [ \partial_{t}+v, \hat{u}_k ] =0   \;.       \label{chsy}
\end{equation}

We shall now switch to this equivalent
representation of symmetries on the flow space $R^3$ because
this form is more suitable to present their geometric features. 
Since the bracket of characteristic vector fields is the same as
the characteristic form of the bracket, the hierarchy (\ref{infsym})
can be written as
$[...[[ \rho^{-1}_{\varphi} B,\hat{u} ],\hat{u} ],...]$ and
generates time-dependent vector fields on $R^3$ satisfying Eqs.(\ref{chsy}).
The Hamiltonian vector fields (\ref{infsym}) are divergence-free
with respect to the Liouville volume. To this end, it would be
appropriate to set $\rho_{\varphi}=1$. This greatly simplifies
the general expressions for symmetries and
allows a better manifestation of their generic properties.

Using notations of three dimensional vector calculus
and the divergence-free property of ${\bf B},\hat{\bf u}_1$,
the hierarchy of symmetries (\ref{infsym}) in their
characteristic form can be obtained recursively from
\begin{equation}
  \hat{\bf u}_{k}=(\nabla \times \hat{\bf u}_1 \times)^k {\bf B}
  =\nabla \times ( \hat{\bf u}_{1} \times \hat{\bf u}_{k-1}) \;,\;\;\;
  k=2,3,4,...    \label{uuu}
\end{equation}
where the operator $\nabla \times \hat{\bf u}_1 \times$ is the "curl of
cross-product with $\hat{\bf u}_1$". We find
\begin{eqnarray}
 & & \hat{\bf u}_{1} \equiv \hat{\bf u}
              =  \nabla \varphi \times \nabla h \;,\;\;\;\;
  \hat{\bf u}_{2}= \nabla \varphi \times \nabla B(h) \;,\;\;\;\;
              \nonumber  \\
 & &  \hat{\bf u}_{3}= \nabla \varphi \times
   \nabla [ \nabla B(h) \cdot  ( \nabla \varphi \times \nabla h ) ] 
  \;, \dots \;,
  \hat{\bf u}_{k} =  \nabla \varphi \times \nabla
  \hat{\bf u}_{k-1}(h) \;, \dots        \label{hier}
\end{eqnarray}
for the first few and the generic members of the hierarchy.
They are made up of gradients of
conserved functions and this property is preserved
under the action of symmetries.
Hence, their flow lines are 
described by intersections of surfaces defined by conserved
functions $\varphi, \; h, \; {\bf B} \cdot \nabla \varphi , \;
{\bf B} \cdot \nabla h,\; (\hat{\bf u}_1 \cdot \nabla)
({\bf B} \cdot \nabla h),\; ...$
of the velocity field.

The vector fields $\hat{\bf u}_{k},\;k=1,2,3,...$ 
are manifestly divergence-free (cf. Eqs.(\ref{uuu}))
and are in Clebsch form (cf. Eqs.(\ref{hier})). Then, as pointed
out in (Exercise 1.4-1 of) Ref.\cite{mr}, we can introduce vector
potentials ${\bf A}_k$ and, moreover, write $\hat{\bf u}_k$'s as
Hamiltonian vector fields on $R^3$.
From Eqs.(\ref{uuu}) we can read off the potentials as
\begin{equation}
  {\bf A}_{1}= \varphi \nabla h \;,\;\;\;
  {\bf A}_{k}= \hat{\bf u}_{1} \times \hat{\bf u}_{k-1}
  =  (\hat{\bf u}_1 \times \nabla \times)^{k-1} {\bf A} \;,\;\;\;
  k=2,3,4,...    \label{aaa}
\end{equation}
where the operator generating ${\bf A}_k$ from the potential
${\bf A}$ of ${\bf B}$ is to "take curl then multiply with
$\hat{\bf u}_1$".

The Hamiltonian structure of $\hat{u}_k$'s is defined by the Poisson bracket
\begin{equation}
  \{ f,g \}_{\varphi} = \nabla \varphi \cdot \nabla f \times \nabla g
\label{brm}    \end{equation}
which is characterized by the Hamiltonian function $\varphi$
of the suspension (\ref{tvel}).
The Hamiltonian functions for the vector fields (\ref{hier}) are
\begin{equation}
  h_{1}= h \;,\;\;\; h_2= \hat{u}_0(h)\;,\;\;\;
   h_3=- \hat{u}_1 (\hat{u}_0(h))\;,\;\;\; \cdots
\end{equation}
from which it follows by induction that for the symmetry $\hat{u}_k$
\begin{equation}
  h_k=(-1)^k \hat{u}_{1}^{k-2}(\hat{u}_0(h)) \;,\;\;k=3,4,5,...
\end{equation}
is the Hamiltonian function.
These are all conserved functions of the velocity field
which are in the form of potential vorticity. Thus, the bracket
(\ref{brm}) may be interpreted as to define the Poisson bracket algebra
of generalized potential vorticities on the flow space.
We note finally that the bracket (\ref{brm})
is induced from and is compatible with the Poisson bracket (\ref{pobi})
on $R \times R^3$ defined by the symplectic two-form $\Omega$.

\newpage

\section{Invariant differential forms}        \label{invariant}

We shall construct invariant differential forms of the velocity
field associated with the infinitesimal symmetries.
The techniques of constructing such invariants together with
the infinite dimensionality of the symmetry algebra will
result in proliferation of invariant forms.
However, a detailed analysis of them will lead us to identify
a basic hierarchy to formulate the laws
of general vorticity and helicity conservations.

Since the time components $\xi_k$ of symmetries (\ref{uk}) are
conserved functions, the right hand side of the symmetry criterion
(\ref{symm}) vanishes and this makes the three-form
\begin{equation}
  \alpha_k = i(U_k)(\mu) = \xi_k dx \wedge dy \wedge dz
  -(\xi_k {\bf v} + \hat{\bf u}_k) \cdot
  d{\bf x} \wedge d{\bf x} \wedge dt          \label{alfa}
\end{equation}
which is independent of our choice for $\rho_{\varphi}$,
an absolute invariant of $\partial_t+v$. This means,
\begin{equation}
  {\cal L}_{\partial_t+v}(\alpha_k)
  = i([ \partial_t+v,U_k ])(\mu) =0          \label{ukin}
\end{equation}
where we used the identity
\begin{equation}
  {\cal L}_V i(U)- i(U) {\cal L}_V = i([ V,U ])     \label{ident}
\end{equation}
and that $v$ is divergence-free. $\alpha_k$'s are closed via
conservation of $\xi_k$'s.
The commutativity of the Lie derivative with the interior product
for $\partial_t+v$ implies that the two-forms
\begin{equation}
 \Theta_k = i(\partial_t+v)(\alpha_k)
     =- \hat{\bf u}_k \cdot d{\bf x} \wedge d{\bf x}
     -( \hat{\bf u}_k \times {\bf v}) \cdot d{\bf x} \wedge dt
\label{bigt}       \end{equation}
are also absolutely invariant. They are closed by Eq.(\ref{ukin})
and the closure of $\alpha_k$'s or, by direct computation using
Eqs.(\ref{chsy}) and divergence-free property of $\hat{u}_k$'s.
Employing the Poincar\'{e} lemma, we introduce one-forms $\theta_k$
\begin{equation}
 \Theta_k = -d \theta_k \;,\;\;\;
   \theta_k = \psi_k dt + {\bf A}_k \cdot d{\bf x}
   \;,\;\;\; \hat{\bf u}_k = \nabla \times {\bf A}_k     \label{steta}
\end{equation}
and $\psi_k$ are determined from the equations
\begin{equation}
  {\partial {\bf A}_k \over \partial t} +
    \hat{\bf u}_k \times {\bf v} = \nabla \psi_k \;.  \label{psi}
\end{equation}
Commutativity of the Lie derivative with the exterior derivative
and the Poincar\'{e} lemma imply that $\theta_k$'s are
relative invariants. That means, their Lie derivatives with
respect to $\partial_t+v$ are exact differentials
\begin{equation}
  {\cal L}_{\partial_t+v}(\theta_k) =
  di(\partial_t+v)(\theta_k) = d \chi_k  \;,\;\;\;\;
  \chi_k \equiv  \psi_k+{\bf A}_k \cdot {\bf v}          \label{chi}
\end{equation}
which follow from the definitions (\ref{bigt}) and (\ref{steta}).

\subsection{A plethora of invariants}

It can be seen using the identities (\ref{iden}) and (\ref{ident})
that for an invariant form $\alpha$ and an infinitesimal symmetry
$U_l$ of $v$, the Lie derivative ${\cal L}_{U_l}(\alpha)$ and
the interior product $i(U_l)(\alpha)$ are also invariants.
In fact, $\Theta_k$ is obtained from $\alpha_{k}$ by taking interior
product with $\partial_t+v$ which is trivially a symmetry. 

Similarly, we have the absolutely invariant two-forms
$\alpha_{lk}=i(U_l)(\alpha_k)$ obtained from $\alpha_k$'s.
Their Lie derivatives
give $d \alpha_{lk}=i([U_l,U_k])(\mu)$ which may be regarded
as redefinitions of $\alpha_k$'s in the algebra of vector fields
(\ref{infsym}).

From $\Theta_k=\alpha_{0k}$ we get absolute invariants
$\Theta_{lk}=i(U_l)(\Theta_k)$ and
$d \Theta_{lk}=i(\partial_t+v)(d \alpha_{lk})$, the latter of which
have similar interpretations of being redefinitions of $\Theta_k$'s.

The relatively invariant one-forms
$\theta_k$ result in conserved functions
$\theta_{lk}= \xi_l \psi_k + \hat{\bf u}_l \cdot {\bf A}_k$
of the velocity field whenever $\chi_k$ in Eqs.(\ref{chi})
are conserved under the flow of the symmetry $U_l$.
Lie derivatives of $\theta_k$'s add the
differential $-d \theta_{lk}$ to the absolute invariants $\Theta_{lk}$
to produce new invariant one-forms which are also absolute invariants
if $\theta_{lk}$ are conserved functions of $v$.

In addition to taking Lie derivative and interior product with
infinitesimal symmetries, forming wedge products of invariants
also produces new invariants. For example,
$\theta_k \wedge \Theta_l$, $\theta_k \wedge \alpha_l$  and
$\theta_k \wedge \Theta_l \wedge \theta_l$ 
are relative invariants while
$\Theta_k \wedge \Theta_l =0 \; \forall k,l=1,2,3,...$

Thus, the above procedures seem to generate a plethora of invariant
forms for the velocity field. However, we find some of them to be
identical 
\begin{eqnarray}
  \theta_k \wedge  \alpha_l = \theta_{lk} \mu =
  [  \chi_k  \xi_l  + {\bf A}_k  \cdot \hat{\bf u}_l  ]  \mu   \\
  \theta_k \wedge \Theta_l \wedge \theta_l =
      \chi_l ({\bf A}_k  \cdot \hat{\bf u}_l  ]  \mu   
\end{eqnarray}
and that they consist of functions available from more simple invariants.
Moreover, we can get rid of the proliferation of invariants by
imposing physical restrictions and utilizing the Lie algebraic
structure of the infinitesimal symmetries.

We first observe that
the invariants $\alpha_k$ consist of two parts which decouple
the velocity field from its symmetries and hence
are physically not much relevant. To see this we write $\alpha_k$
as the sum of two absolute invariants
\begin{equation}
  \alpha_k =  \xi_k \sigma  - \hat{\bf u}_k \cdot
  d{\bf x} \wedge d{\bf x} \wedge dt  \;,\;\;\;\;    
 \sigma =  \Pi_{i=1}^{3} ( dx^i - v^i dt )           \label{alfas}
\end{equation}
where $\sigma$ is the well-known invariant three-form that appears
in the formulation of continuity equation and has no effect
in the definition of $\Theta_k$'s.

The invariants $\alpha_{lk}$, on the other hand, have a similar
structure as those of $\Theta_l$ involving appropriate
differences of the symmetries $\hat{u}_l$ and $\hat{u}_k$
and thus can be taken to be equivalent.
Likewise, due to the Lie bracket relations between symmetries 
we can regard $d \alpha_{lk}$ and $d \Theta_{lk}$
to be equivalent to (\ref{alfa}) and (\ref{bigt}), respectively.

\section{General conservation laws}

This analysis of invariants leads us to conclude that appropriate
expressions of the laws of general vorticity and helicity
conservations may be
best given in terms of the basic hierarchy consisting of
$\Theta_k$ and $\theta_k$.
We shall formulate the conservation laws as a restatement of
infinitesimal conditions on these invariants in terms of
integration over submanifolds of the flow domain.
\begin{theorem}
Let $S_t$ and $D_t$ be two and three dimensional regions in
$R \times R^3$ advected by the graph $(t, g_t)$ of the flow
of the velocity field $v$.
Denote their smooth boundaries by $\partial S_t$ and $\partial D_t$.
Then,

(1) The exact two-forms $\Theta_k=-d \theta_k$ are the basic
ingredients of
general vorticity conservation law which can be expressed as
\begin{equation}
 {d \over dt}  \int_{S_t} \Theta_k =
     - {d \over dt}  \oint_{\partial S_t} \theta_k =
      -  \oint_{\partial S_t} d \chi_k  \equiv   0
 \;,\;\;\; k=0,1,2,...   \label{gvc}
\end{equation}
where $\Theta_0$ is the symplectic two-form $\Omega$.

(2) The relatively invariant three-forms $\theta_k \wedge d \theta_l$
define the general helicity conservation laws
\begin{equation}
 {d \over dt}  \int_{D_t} \theta_k \wedge d \theta_l =
 \oint_{\partial D_t}  \chi_k  d \theta_l
   = - \oint_{\partial D_t} d \chi_k \wedge  \theta_l =0   \label{gec}
\end{equation}
if either $\hat{u}_l$ is tangent to the projection onto spatial
domain of $\partial D_t$
or, $\chi_k \equiv i(\partial_t+v)(\theta_k)=constant$.

(3) If, on the other hand, $\hat{u}_l$ is tangent to the surfaces
$\chi_k=constant$ then the generalized helicity densities in
$\theta_k \wedge d \theta_l$ are Lagrangian invariants, that is,
conserved functions of the velocity field.
Moreover, they belong to the Poisson bracket algebra (\ref{brm})
on $R^3$ of generalized potential vorticities.        \label{gcl}
\end{theorem}
For an equivalent formulation, using the one-forms $\theta_k$,
of the first part of proposition(\ref{gcl})
one replaces exactness and absolute
invariance with Eqs.(\ref{psi}) and relative invariance, respectively.
In obtaining the second part, we used the properties of the velocity field
that characterize it as the generator of volume preserving
diffeomorphisms. Namely, it is divergence-free and is tangential to the
boundary of $D_t$. Eqs.(\ref{gvc}) and the condition that
the helicity is conserved
for $\chi_k =constant$ follow from the fact that the integral of
exact forms over boundaries is identically
($\partial \circ \partial \; \cdot \equiv$) zero
via Stokes' theorem. Thus, in the latter case, there will be no boundary
terms contributed by $\hat{\bf u}_l$,
meaning that the three-forms are absolutely invariant
and they result in the last conclusion of the proposition.
To complete the proof of proposition(\ref{gcl})
we shall supply, in the next section, coordinate expressions for
these geometric arguments as well as for the global
formalism of conservation laws.

We end this section with a remark on the boundaries of the
integration domains $S_t$ and $D_t$ which are two and three
dimensional subsets of the time-extended space $R \times R^3$. 
For such decomposition of space-time, the one and two
dimensional boundaries $\partial S_t$ and $\partial D_t$ 
do not close up, but rather have
extention along a closed, finite interval of the time axis.
Their projection onto spatial domain are a closed curve and
a closed surface in the usual sense. For example,
a circular helix whose finite-lenght {\it axis} is directed along
the time axis is the boundary of a two-surface and thus is a
closed curve in $R \times R^2$.

\section{Demonstrations}

We shall give explicit coordinate expressions for the general
conservation laws of proposition(3).
In the particular case of the Euler flow
we shall show that the dynamical equations themselves,
Bernoulli's equation, the Kelvin's circulation theorem,
the Helmholtz' vorticity theorem, conservations
of helicity and potential vorticity
can be obtained from the invariance conditions of the basic
hierarchy $\Theta_k$, $\theta_k$ and from the global formulations
(\ref{gvc}) and (\ref{gec}).

\subsection{Bernoulli's equation}

For ideal fluids and for $k=0$, we consider the invariance condition
(\ref{chi}) of the canonical one-form in Eq.(\ref{cone2}) with
$\psi$ as given in Eqs.(\ref{psio}). Using the rotational form
\begin{equation}
{\partial {\bf v} \over \partial t} 
 - {\bf v} \times {\bf w} =
  - \nabla ( {P \over \rho} + {1 \over 2} v^2 )     \label{reuler}
\end{equation}
of the Euler equations we obtain from Eq.(\ref{chi})
the Bernoulli's equation
\begin{equation}
 {\partial \over \partial t} ({1 \over 2} v^2)
 + {\bf v} \cdot \nabla ({P \over \rho} + {1 \over 2} v^2) =0
\label{bern}   \end{equation}
where $P + \rho v^2 /2 $ is the total (or stagnation) pressure.
This says that the non-invariance of the kinetic energy
under the flow of velocity field is due to the existence of
pressure.

Although, Eq.(\ref{bern}) is a direct consequence of (\ref{reuler}),
our purpose to include this and the next example is to demonstrate
the interplay between the present geometric framework for the
Lagrangian description and the Eulerian dynamical equations.

\subsection{The Euler equations}

The hierarchy of Eqs.(\ref{gvc}) contains, in particular, the
dynamical equations for the Euler flow. This follows from the
relative invariance of the one-forms $\theta_k$. 
Eq.(\ref{chi}) for $k=0$ can be solved for the Lie derivative
\begin{equation}
  {\cal L}_{\partial_t+v}({\bf v} \cdot d{\bf x}) =
  {\partial \over \partial t} ({1 \over 2} v^2) dt
     + \nabla ({1 \over 2} v^2 -{P \over \rho}) \cdot d {\bf x}  
\end{equation}
where we used Bernoulli's equation to simplify the right hand side.
Note that the Lie derivative is the one on $R \times R^3$.
In terms of the differential operators on $R^3$
this expression reduces to
\begin{equation}
   {\partial \over \partial t}    ({\bf v} \cdot d{\bf x})+
  {\cal L}_{v}({\bf v} \cdot d{\bf x}) =
    d ({1 \over 2} v^2 -{P \over \rho})   \label{ineu}
\end{equation}
which is the invariant form of the Euler equations of ideal fluid.
It is this form that can be generalized for flows on general
Riemannian manifolds of arbitrary dimensions \cite{mr}.

\subsection{Helmholtz' vorticity theorems}

We first remark that in the integration of the two-forms $\Theta_k$ over
two-surfaces advected by the fluid motion, the term
$( \hat{\bf u}_k \times {\bf v}) \cdot d{\bf x} \wedge dt$
has no contribution because its pull-back to Lagrangian
coordinates by the solution $g_t$ of (\ref{vel}) vanishes.
Then, (\ref{gvc}) becomes an expression for the general form of
Helmholtz' vorticity theorem
\begin{equation}
 {d \over dt}  \int_{S_t} \Theta_k =
   {d \over dt}  \int_{S_t}
   \hat{\bf u}_k \cdot d{\bf x} \wedge d{\bf x} =0      \label{helm}
\end{equation}
which reduces for $k=0$, $\hat{\bf u}_0 = {\bf w} =
\nabla \times {\bf u}$ to the conservation of vorticity flux
and, for $\hat{\bf u}_0 = {\bf B}$ to the Alfv\'{e}n's theorem
of dynamo theory.

\subsection{Kelvin's circulation theorems}

For two-surfaces $S_t$ with smooth boundary $\partial S_t$
the invariance of the two-forms $\Theta_k$ gives
\begin{equation}
  0 =  {d \over dt}   \oint_{\partial S_t} \theta_k
  = {d \over dt}   \oint_{\partial S_t} {\bf A}_k  \cdot d{\bf x}
\label{kel}      \end{equation}
via Stokes' theorem and the generic form of symmetry generators
as curl vectors.
This is a generalization of the Kelvin's circulation theorem
for which the usual form is obtained when $k=0$.
It is the conservation of circulation of the velocity
field ${\bf A}_0={\bf v}$ for the Euler flow and of
the vector potential ${\bf A}_0 = {\bf A}$ for magnetization problems.

\subsection{General vorticity conservations}       \label{sgvc}

An immediate generalization to the infinite dimensional left Lie
algebra of symmetries of the well-known equivalence
of the Helmholtz' and the Kelvin's theorems \cite{mr}
follows directly from Eqs.(\ref{gvc}), (\ref{helm}) and (\ref{kel}).
For two-surfaces
$S_t=g_t(S)$ with smooth boundary $\partial S_t$ we obtain
\begin{equation}
 \int_{S} g_t^*(\hat{\bf u}_k \cdot d{\bf x} \wedge d{\bf x})
 = \int_{S_t} \hat{\bf u}_k \cdot d{\bf x} \wedge d{\bf x}
 = \oint_{\partial S_t} {\bf A}_k \cdot d{\bf x} = \; constant
\end{equation}
where $g_t$ is a flow of the velocity field.
The constancy of the first integral is the preservation of
coadjoint orbits of $\Theta_k$'s,
of the second is the Helmholtz' theorem on conservation
of flux of flow lines of the infinitesimal symmetry $\hat{\bf u}_k$,
and of the third is the Kelvin's theorem on conservation
of circulation of vector potential ${\bf A}_k$ for  $\hat{\bf u}_k$.
The fact that, for fixed $\hat{\bf u}_k$, these expressions hold
for all solutions $g_t$ of the velocity field
is a manifestation of the right invariance on the cotangent
bundle of the group of volume preserving diffeomorphisms.

\subsection{General helicity conservations}          \label{ghc}

Eulerian invariants
are conservation laws of dynamical equations (\ref{geuler}) which are
divergence expressions of the form
\begin{equation}
     {\partial T  \over \partial t} + \nabla \cdot {\bf P} = 0  \label{cola} 
\end{equation}     
where the conserved density $T$ and the flux ${\bf P}$ are functions
of $t,{\bf x}$, the Eulerian fields and their derivatives.
Observing that Eq.(\ref{cola}) can be expressed as the closure of
the three-form
\begin{equation}
   - T dx \wedge dy \wedge dz +
    {\bf P} \cdot d{\bf x} \wedge d{\bf x} \wedge dt
\end{equation}
we shall derive the Eulerian conservation laws as secondary
invariants obtained from the basic hierarchy (\ref{bigt}) and
(\ref{steta}).
These are the relative invariants $\theta_k \wedge d \theta_l$
which are closed via Eqs.(\ref{chsy}) and (\ref{psi}).
The corresponding conservation laws are expressed by
Eqs.(\ref{gec}) of proposition(\ref{gcl}).

The case $k=l=0$ is exceptional because $\Omega$ is non-degenerate.
In this case, the identity
\begin{equation}
 d( \theta \wedge \Omega) + \Omega \wedge \Omega =0   \label{s3}
\end{equation}
gives the conservation of usual (magnetic) helicity
\begin{equation}
 { \partial \over \partial t} ( {\bf A} \cdot {\bf B} )
       + \nabla \cdot ( ( {\bf A} \cdot  {\bf B}) {\bf v} 
  - (\psi + \varphi + {\bf v} \cdot  {\bf A}) {\bf B} ) = 0    \label{hel}
\end{equation}
at which a hierarchy of others is anchored. For $k,l \neq 0$ we find
\begin{equation}
 {\partial \over \partial t}
          ( {\bf A}_k \cdot \nabla \times {\bf A}_l)+
  \nabla \cdot ( {\bf A}_k \times ({\bf v} \times
  (\nabla \times {\bf A}_l))
     - \psi_k \nabla \times {\bf A}_l )=0      \label{hel}
\end{equation}
associated with a pair of symmetries. Using the identity
\begin{equation}
 {\bf A}_k \times ( {\bf v} \times \hat{\bf u}_l ) =
       ( \hat{\bf u}_l \cdot {\bf A}_k) {\bf v}
       - ( {\bf v} \cdot {\bf A}_k) \hat{\bf u}_l      \label{bac}
\end{equation}
and the divergence-free properties of ${\bf v}$, $\hat{\bf u}_l$
we can write Eq.(\ref{hel}) as
\begin{equation}
 {\partial \over \partial t}
     ( {\bf A}_k \cdot \hat{\bf u}_l)+
  {\bf v} \cdot \nabla ( {\bf A}_k \cdot \hat{\bf u}_l)
   - \hat{\bf u}_l \cdot  \nabla  \chi_k =0      \label{hell}
\end{equation}
from which one can immediately conclude the results in the second and
the third items of proposition(3).
Thus, the functions
\begin{eqnarray}
  {\bf A} \cdot \nabla \times {\bf A} \;,\;\;\;
  {\bf A} \cdot \nabla \varphi \times \nabla h \;,\;\;\;
  \varphi {\bf B} \cdot \nabla h \;,\;\;\;   \nonumber \\
 (\hat{\bf u}_1 \times \nabla \times)^{k-1} {\bf A} \cdot
 \nabla \times (\hat{\bf u}_1 \times \nabla \times)^{l-1} {\bf A} 
 \;,\;\;\;\;k,l=2,3,...  \label{helden}      \end{eqnarray}
are the first few and the generic members of the conserved densities
associated to the general helicity conservation (\ref{gec}).
The usual helicity and the potential vorticity
are included as the first and the third elements of this hierarchy.

The conservation of potential vorticity can also be
obtained from the absolute invariance of the four-form
$\Omega \wedge \Omega$ by integration over a region in $R \times R^3$.
This, in turn, implies the constancy of the integral of the
three-form $\theta \wedge \Omega$ over closed three dimensional
domains in $R \times R^3$. 
For $k=l=1,2,3,...$ the densities (\ref{helden}) vanish because the
symmetries $\hat{\bf u}_{k}$ have Clebsch representations.

\subsection{Topological interpretations}

In the particular case of $l=0$ and $k=0,1,2,...$, the hierarchy
$\hat{\bf u}_0 \cdot {\bf A}_k$ of invariants is related to the linking
number of trajectories of vorticity $\hat{\bf u}_0 = {\bf w}$
or magnetic $\hat{\bf u}_0 = {\bf B}$ fields
with those of symmetries \cite{mofat}-\cite{khch}.
For $k=0$, the integral of this function coincides 
either with the self-linking number of vorticity or magnetic field,
which is also called helicity, or, with the linking number of
${\bf w}$ and ${\bf B}$ expressed by the integrals of
${\bf w} \cdot {\bf A}$ or
${\bf v} \cdot {\bf B}$. This latter result, when first proven
with the second integrand in
\cite{khch} for ideal magnetohydrodynamic equations, was interpreted
to be unexpected because ${\bf w}$ is not frozen. Here, we see
that it is sufficient to have ${\bf B}$, and in general $\hat{\bf u}_k$,
to be frozen.

The closed two-forms $\Theta_k$ can be associated with two-dimensional
foliations of four-space by the integral curves of $U_k$ and $V$.
The conditions $\Theta_k \wedge \Theta_l=0$ means that intersection
of leaves of any two such foliations is a one-dimensional
foliation. The leaves of the latter are trajectories of the
suspended velocity field. Moreover, for any $k,l$, $(V,U_k,U_l)$
form a three-dimensional foliation. In this case, the integrals
of the relative invariants $\theta_l \wedge \theta_k \wedge \Theta_k$
are shown to be related to the average linking number of the
foliation of $\Theta_k$ with the vector field $U_{kl}$ defined
by $i(U_{kl})(\mu)=d(\theta_k \wedge \theta_l)$ \cite{spn},\cite{arsin}.
The characteristic form of this is the vector field
$\hat{\bf u}_{kl} = \chi_k \hat{\bf u}_l- \chi_l \hat{\bf u}_k$
on $R^3$.

We conclude from these examples that it is the hierarchy of
secondary and higher order invariants of
particle relabelling symmetry which contains topological
informations about the flow domain.

\subsection{Lagrangian invariants}       \label{lcl}

Lagrangian invariants are functions which are conserved under
the flow of the velocity field (\ref{vel}).
The Eulerian conservation laws (\ref{hell}) reduce to the invariance
of the functions $\hat{\bf u}_k \cdot {\bf A}_l$, $k,l \neq 0,1$
in the Lagrangian description when the condition
$\hat{\bf u}_k \cdot \nabla \chi_l=0$ is satisfied.
For the special case of $l=0$ for which
$ \chi_0 = \psi + \varphi + {\bf v} \cdot {\bf A}$,
this can be verified using Eqs.(\ref{aeq}) and the identity
${\bf v} \times (\nabla \times {\bf A}) + ({\bf v} \cdot \nabla) {\bf A}
= v^i \nabla A_i$.
Similarly, one obtains from Eq.(\ref{geuler})
and the symmetry condition for $\hat{\bf u}_k$ that
$\hat{\bf u}_k \cdot {\bf v}$ is a Lagrangian invariant if
\begin{equation}
   \hat{\bf u}_k \cdot ( {\bf F} + {1 \over 2} \nabla v^{2})=0 \;
   ,\;\;\; k=0,1,2,...            \label{invu}
\end{equation}
is satisfied by a general force field ${\bf F}$.
It can be checked that for the Euler equations of ideal fluid
this coincides with the general criterion.

The hierarchy of Lagrangian invariants includes the helicities
${\bf v} \cdot {\bf w}$, ${\bf A} \cdot {\bf B}$, the potential
vorticities ${\bf w} \cdot \nabla h$, ${\bf B} \cdot \nabla h$
as well as the invariants ${\bf w} \cdot {\bf A}$ and
${\bf v} \cdot {\bf B}$ of magnetization problems and, furthermore
${\bf v} \cdot \nabla \varphi \times \nabla h$,
${\bf v} \cdot \nabla \varphi \times \nabla h$ which are obtained
for the first two values of $k,l$.
It can be seen using the Hamiltonian form with (\ref{brm}) of 
$\hat{u}_k$'s that the Lagrangian invariants
$\hat{\bf u}_k \cdot {\bf A}_l$ belong to the Poisson
bracket algebra (\ref{brm}) of generalized potential
vorticities $h_k$ and can be obtained from the Poisson
brackets of functions $h_k$ and $h_l$.

An immediate application of these conserved functions of the
velocity field may be the construction of volume preserving
diffeomorphisms which are also automorhisms of the symplectic
two-form (\ref{symp2}). This can be achieved by choosing
a set $y^i(t,{\bf x}) \;,\;i=1,2,3$ of functionally independent
Lagrangian invariants which further satisfy the condition
\begin{equation}
  det \left( {\partial y^i \over \partial x^j} \right)
  = \nabla y^1 \cdot \nabla y^2 \times \nabla y^3 =1     \label{det}
\end{equation}
for volume preservation. In particular, if we let $y^1= \varphi$
then $\nabla y^2 \times \nabla y^3$ must be ${\bf B}$ (recall that
we set $\rho_{\varphi}=-B(\varphi)=1$). Thus, the other two Lagrangian
invariants $y^2$ and $y^3$ are the Clebsch variables $p,q$ for ${\bf A}$
(cf. Eqs.(\ref{aeq})).
From proposition(1) we conclude that this particular diffeomorphism
is the one which brings the symplectic two-form $\Omega$ into canonical form.
It follows from Eq.(\ref{det}) and (\ref{brm}) that
\begin{equation}
       \{ q,p  \}_{\varphi} = 1 \;.
\end{equation}

\section{Discussions and conclusions}

We presented a geometric approach to the explicit
construction of symmetries starting from an
initial one and utilizing a symplectic structure both of which
are implicit in the Eulerian equations constraining the velocity field.
We may conclude that,
as the simplicity and the Lie-Poisson structure
of Eulerian description derives from the particle
relabelling symmetry \cite{via}-\cite{salmon}, the Eulerian description,
in its turn, enables us to construct the Lie algebra of these symmetries.

Therefore, depending on the structure of the Eulerian
dynamical equations, the
present construction may be generalized to other hydrodynamic systems.
We demonstrated that the Euler flow in arbitrary
Riemannian manifolds allows this sort of generalizations.
A treatment for compressible flows may be based on the
recognition that the so-called exponential vector field \cite{olver}
$exp( \int \nabla \cdot {\bf v} dt ) {\bf w}$ is a time-dependent
infinitesimal symmetry of the velocity field.
We refer to Refs.\cite{khch} and \cite{arsin} for other examples
of hydrodynamic systems which admit a vectorial frozen-in field
and hence can be analysed with the present geometric framework.

It would be reasonable to compare the overall picture for invariants of 
three dimensional flow with that of two dimensional motion
because, although, there is no essential difference in their group
theoretical descriptions, they have been observed to have some
different quantitative aspects
\cite{arsin}-\cite{gk},\cite{khch},\cite{via},\cite{mr}.
We showed that
some properties of two dimensional flows such as the representation
of area preserving diffeomorphisms as Hamiltonian vector fields
and the existence of infinitely many enstrophy type integrals,
can be realized, with appropriate modifications of geometric tools,
for the three dimensional flows as well.

Moreover, due to dimensional reasons we obtained a much richer
structure of invariants for the present case.
Namely, the geometric framework at our disposal led us to
discriminate the invariants into three classes each having
a certain degree of relation to the particle relabelling
symmetry. Thus, for three dimensional flows,
in addition to general vorticity conservation, we can speak of
a general helicity conservation and moreover, under certain
conditions, of general Lagrangian invariants.

The geometric character and the algebraic consequences
of our construction and classification of
symmetries and invariants distinguish the present work
from the approaches that employ the analysis of defining
equations as the main tool \cite{hen},\cite{sagdeev},\cite{kuzmin}.
Such constructions require that
the algebraic structure of the solution set
consisting of symmetries and invariants be separately treated. Instead,
we gave a recursive construction of symmetries via the Jacobi identity
and, moreover, realized them as the Hamiltonian
automorphisms of the symplectic two-form. This
ensures that the infinitesimal symmetries thus obtained
constitute a Lie algebra
because, although not every infinite set of vector fields form
a Lie algebra, Hamiltonian vector fields do so.

\end{document}